\documentclass[12pt,preprint]{aastex}

\begin{document}

\title{Collapse and Fragmentation of Molecular Cloud Cores. X. 
Magnetic Braking of Prolate and Oblate Cores.}

\author{Alan P.~Boss}
\affil{Department of Terrestrial Magnetism, Carnegie Institution of 
Washington, 5241 Broad Branch Road, NW, Washington, DC 20015-1305}
\email{boss@dtm.ciw.edu}

\begin{abstract}

 The collapse and fragmentation of initially prolate and oblate, 
magnetic molecular clouds is calculated in three dimensions with a 
gravitational, radiative hydrodynamics code. The code includes 
magnetic field effects in an approximate manner: magnetic pressure, 
tension, braking, and ambipolar diffusion are all modelled. The 
parameters varied for both the initially prolate and oblate clouds
are the initial degree of central concentration of the radial density
profile, the initial angular velocity, and the efficiency of magnetic
braking (represented by a factor $f_{mb} = 10^{-4}$ or $10^{-3}$).
The oblate cores all collapse to form rings that might 
be susceptible to fragmentation into multiple systems. The outcome of 
the collapse of the prolate cores depends strongly on the initial density
profile. Prolate cores with central densities 20 times higher
than their boundary densities collapse and fragment into
binary or quadruple systems, whereas cores with central densities
100 times higher collapse to form single protostars embedded
in bars. The inclusion of magnetic braking is able to stifle
protostellar fragmentation in the latter set of models, as 
when identical models were calculated without magnetic braking
(Boss 2002), those cores fragmented into binary protostars.
These models demonstrate the importance of including magnetic
fields in studies of protostellar collapse and fragmentation,
and suggest that even when magnetic fields are included, 
fragmentation into binary and multiple systems remains as a 
possible outcome of protostellar collapse.

\end{abstract}


\section{Introduction}

 Fragmentation during protostellar collapse is widely accepted
to be the primary mechanism for the formation of binary and multiple 
star systems (e.g., Lafreni\`ere et al. 2008; Chen et al. 2008).
While it is clear that the overall form of the initial mass function
for stars is directly tied to the initial conditions for protostellar
collapse, i.e., the mass function of dense cloud cores (e.g.,
Dib et al. 2008; Swift \& Williams 2008), fragmentation is necessary
for producing binary star systems within these individual dense cores
(Lafreni\`ere et al. 2008; Chen et al. 2008).

 Three dimensional calculations of the collapse of centrally condensed,
rotating cloud cores have been computed for quite some time 
(e.g., Boss 1993) and continue to attract theoretical attention
(e.g., Saigo et al. 2008; Machida 2008; Commercon et al. 2008).
These calculations neglected the effects of magnetic fields. However, 
observations of OH Zeeman splitting in dark cloud cores have shown that 
magnetic fields are often an important contributer to cloud support 
against collapse for densities in the range of $10^3 -10^4$ cm$^{-3}$
(Troland \& Crutcher 2008). Given this observational constraint,
it is clear that three dimensional hydrodynamical collapse 
calculations should include magnetic field effects as well as 
self-gravity and radiative transfer (e.g., Boss 1997, 1999, 2002, 
2005, 2007). Magnetic fields are now being included in other 
three dimensional collapse models as well (e.g., 
Machida et al. 2004, 2005a,b, 2007, 2008; Kudoh et al. 2007;
Price \& Bate 2007, 2008). In particular, Price \& Bate (2007) 
found that while magnetic pressure acts to resist fragmentation
during collapse, magnetic tension can actually aid fragmentation,
confirming the results found by Boss (2002). Machida et al. (2004,
2005a,b, 2007, 2008) generally found that binary fragmentation 
could still occur provided that the initial magnetic cloud core
rotated fast enough.

 Magnetic braking is effective at reducing cloud rotation 
rates during the pre-collapse cloud phase, but has relatively
little effect during the collapse phase, according to the two
dimensional magnetohydrodynamics models of 
Basu \& Mouschovias (1994, 1995a,b). However, Hosking \& Whitworth 
(2004) found that rotationally-driven fragmentation could be 
halted by magnetic braking during the collapse phase. Boss (2004)
argued that the thermodynamical treatment employed by 
Hosking \& Whitworth (2004) could have been more important for
stifling fragmentation than magnetic braking, but did not
offer any models of magnetic braking to support this assertion.

 Price \& Bate (2007) presented models of the collapse of
magnetic cloud cores, finding that magnetic pressure was more
important for inhibiting fragmentation than either magnetic
tension or braking, contrary to the results presented by 
Hosking \& Whitworth (2004) and Fromang et al. (2006), who
found no evidence at all for the fragmentation of magnetic clouds.
Fromang et al. (2006) assumed ideal magnetohydrodynamics (MHD), 
i.e., without ambipolar diffusion, a fact that is likely to stifle
fragmentation, whereas Hosking \& Whitworh (2004), Boss (2004),
and Price \& Bate (2007) included ambipolar diffusion. However,
subsequent ideal MHD collapse calculations by Hennebelle \&
Fromang (2008) and Hennebelle \& Teyssier (2008) found that
magnetic clouds could fragment if the initial density perturbation 
was large enough, and they speculated on what would happen if
ambipolar diffusion was included in their models. These fundamental
differences in the results of magnetic cloud collapse calculations
results highlight the need to compare models where only one parameter 
at a time is being changed, so that the true effect of changing
that one parameter can be discerned. Such a comparison is the
major goal of the present study.

 Banerjee \& Pudritz (2006) considered the collapse of magnetized
cloud cores, finding that even though considerable angular momentum
was lost from the cloud core by magnetic outflows from the disk's
surface, the cloud was still able to collapse and fragment into
a close binary protostar system.
 
 Price \& Bate (2007) and  Hennebelle \& Teyssier (2008) both
considered the collapse of spherical, magnetic cloud cores with 
initially uniform density and uniform magnetic field strengths.
Machida et al. (2004, 2005a,b) studied the collapse of initially
cylindrical cloud cores in hydrostatic equilibrium. 
Machida et al. (2008) considered the collapse of cloud cores
with density profiles appropriate for Bonnor-Ebert isothermal
spheres, similar to the Gaussian radial density profile
clouds studied by Boss (1997, 2002) and by this paper. Such
centrally-condensed density profiles represent the best guesses
as to the radial structure of pre-collapse molecular cloud cores 
(e.g., Myers et al. 1991; Ward-Thompson, Motte, \& Andr\'e 1999).

 Boss (2002) modelled the collapse of initially prolate and oblate cores, 
including several magnetic field effects, but ignoring magnetic braking.
Prolate and oblate cloud shapes have been inferred from observations
of suspected pre-collapse molecular cloud cores (e.g., Jones, Basu,
\& Dubinski 2001; Curry \& Stahler 2001). 
Here we use the magnetic braking approximation developed by
Boss (2007), originally applied to filamentary clouds, to examine
the importance of magnetic braking for the same prolate and oblate 
models as those calculated by Boss (2002), allowing a direct comparison
between identical protostellar collapse models with and without
magnetic braking. These models thus directly address the different
outcomes of the models with magnetic braking but without detailed
thermodynamics by Hosking \& Whitworth (2004) compared to those
without magnetic braking but with detailed thermodynamics of Boss 
(2004): the present models include both effects.

\section{Numerical Methods}

 The numerical models are calculated with a three-dimensional hydrodynamics 
code that calculates finite-difference solutions of the equations of 
radiative transfer, hydrodynamics, and gravitation for a compressible 
fluid (Boss \& Myhill 1992). The hydrodynamic equations are solved in 
conservation law form on a contracting spherical coordinate grid,
subject to constant volume boundary conditions on the spherical boundary.
The code is second-order-accurate in both space and time, with the van 
Leer-type hydrodynamical fluxes having been modified to improve stability 
(Boss 1997). Artificial viscosity is not employed. Radiative transfer is 
handled in the Eddington approximation, including detailed equations of 
state and dust grain opacities (e.g., Pollack et al. 1994). The 
code has been tested on a variety of test problems (Boss \& Myhill 1992; 
Myhill \& Boss 1993).

 The Poisson equation for the cloud's gravitational potential is solved by a 
spherical harmonic expansion ($Y_{lm}$) including terms up to $N_{lm} = 32$.
The computational grid consists of a spherical coordinate grid 
with $N_r = 200$, $N_{\theta} = 22$ for $\pi/2 \ge \theta > 0$
(symmetry through the midplane is assumed for $\pi \ge \theta > \pi/2$),
and $N_{\phi} = 256$ for $2 \pi \ge \phi \ge 0$, i.e., with no symmetry
assumed in $\phi$. The radial grid contracts to follow the collapsing inner 
regions and to provide sufficient spatial resolution to ensure satisfaction 
of the four Jeans conditions for a spherical coordinate grid (Truelove et al. 
1997; Boss et al. 2000). The innermost 50 radial grid points are kept uniformly
spaced during grid contraction, while the outermost 150 are non-uniformly 
spaced, in order to provide an inner region with uniform spatial resolution
in the radial coordinate. The $\phi$ grid is uniformly spaced, whereas 
the $\theta$ grid is compressed toward the midplane, where the minimum 
grid spacing is $0.3$ degrees.

\section{Initial Conditions}

 Tables 1 and 2 list the initial conditions for the models. The 
initial models have Gaussian radial density profiles (Boss 1997) 
of the form

\begin{equation}
\rho_i(x,y,z) = \rho_o \ exp\biggl(
- \biggl( {x \over r_a} \biggr)^2
- \biggl( {y \over r_b} \biggr)^2
- \biggl( {z \over r_c} \biggr)^2
\biggr),
\end{equation}

\noindent 
where $\rho_0 = 2.0 \times 10^{-18}$ g cm$^{-3}$ is the initial central 
density. The prolate clouds with central densities 20 times higher than a 
reference boundary density have $r_a = 1.16 R$ and $r_b = r_c = 0.580 R$,
where $R$ is the cloud radius, yielding a axis ratio of 2 to 1.
The oblate clouds with the same degree of central concentration
have $r_a = r_b = 1.16 R$ and $r_c = 0.580 R$. For 100 to 1
density contrasts, the prolate clouds have
$r_a = 0.932 R$ and $r_b = r_c = 0.466 R$, while the 
oblate clouds have $r_a = r_b = 0.932 R$ and $r_c = 0.466 R$. 
Random numbers ($ran(x,y,z)$) in the range [0,1] are used
to add noise to these initial density distributions by multiplying
$\rho_i$ from the above equation by the factor $[1 + 0.1 \ ran(x,y,z)]$.
The cloud radius is $R = 1.0 \times 10^{17}$ cm $\approx$ 0.032 pc 
for all models. 

 The cloud masses are 1.5 $M_\odot$ and 2.1 $M_\odot$, 
respectively, for the prolate and oblate clouds with a 20:1 density 
ratio, and 0.96 $M_\odot$ and 1.5 $M_\odot$, for the prolate and 
oblate clouds with a 100:1 density ratio. With an initial temperature
of 10 K, the initial ratio of thermal to gravitational energy is
$\alpha_i = 0.39$ for the prolate clouds and 0.30 for the oblate clouds
with the 20:1 initial density ratio, while $\alpha_i = 0.55$ for the
prolate clouds and 0.39 for the oblate clouds with the 100:1 initial 
density ratio.                                                     

 Solid body rotation is assumed, with the angular velocity about 
the $\hat z$ axis (short axis) taken to be $\Omega_i$ = $10^{-14}$, 
$3.2 \times 10^{-14}$, or $10^{-13}$ rad s$^{-1}$. These choices of 
$\Omega_i$ result in initial ratios of rotational to gravitational 
energy varying from $\beta_i = 1.2 \times 10^{-4}$ to 0.015 for the 
prolate clouds and $\beta_i = 1.1 \times 10^{-4}$ to 0.013 for 
the oblate clouds. These choices are all consistent with observational 
constraints on the densities, shapes, and rotation rates
of dense cloud cores (e.g., Myers et al. 1991; Goodman et al. 1993;
Ward-Thompson, Motte, \& Andr\'e 1999; Jones, Basu, \& Dubinski 2001; 
Curry and Stahler 2001).

 As in the previous three-dimensional models (Boss 1997, 1999, 2002, 
2005, 2007), the effects of magnetic fields are approximated 
through the use of several simplifying approximations regarding
magnetic pressure, tension, braking, and ambipolar diffusion 
(see Boss 2007 for a derivation of these approximations).
All models assumed an ambipolar diffusion time scale 
$t_{ad}$ = 10 $t_{ff}$, where the free fall time 
$t_{ff} = (3\pi/32 G \rho_0)^{1/2} = 4.7 \times 10^4$ yr for a
central density $\rho_0 = 2.0 \times 10^{-18}$ g cm$^{-3}$. 
The reference magnetic field strength is assumed to be 
$B_{oi} = 200 \mu$G. The magnetic braking factor $f_{mb}$ is 
taken to be either 0.0001 or 0.001. Based on the models
of Basu \& Mouschovias (1994), Boss (2007) estimated that
$f_{mb} \sim 0.0001$. The models with $f_{mb} = 0.001$ are
thus intended to attempt to overestimate the effects of
magnetic braking.

 Note, however, that the calculations of Basu \& Mouschovias
(1994) stopped at central densities of $3 \times 10^9$ cm$^{-3}$,
i.e., before the clouds became optically thick, whereas the
present models are continued well in the optically thick regime.
Hence the implicit assumption that the trends found in the Basu \&
Mouschovias (1994) models and used by Boss (2007) to derive 
the $f_{mb}$ approximation should continue indefinitely may
not be warranted, though the trends do persist over the previous
six orders of magnitude increase in central density of their
models (see Figure 7 of Basu \& Mouschovias 1994). The magnetic
braking studied by Basu \& Mouschovias (1994) is similar to
the magnetic braking produced by disk jets and outflows in the
models by Banerjee \& Pudritz (2006), Hennebelle \& Fromang (2008),
and Hennebelle \& Teyssier (2008). A superior treatment of
magnetic braking beyond the $f_{mb}$ approximation of Boss (2007)
will require a true MHD code.

 With the field strength $B_{oi} = 200 \mu$G, the prolate or 
oblate clouds with 20:1 density contrasts have initial ratios 
of magnetic to gravitational energy of $\gamma_i = 0.58$ or 0.43,
respectively. For density ratios of 100:1, $\gamma_i = 0.81$ for
the prolate clouds and 0.57 for the oblate clouds. The mass to 
flux ratio of these clouds is less than the critical mass to 
flux ratio, making all the clouds formally magnetostatically
stable and hence magnetically subcritical. Protostellar collapse 
cannot occur until ambipolar diffusion leads to sufficient loss 
of magnetic field support for sustained contraction to begin.

\section{Results}

 Tables 1 and 2 list the initial conditions as well as the basic 
outcome of each model, namely the final time to which the cloud was
advanced $t_f$ (in units of the initial cloud free fall time) and
whether the cloud collapsed to form a quadruple system (Q), binary (B),
binary-bar (BB), single-bar (SB), ring (R), or did
not collapse (NC). For convenience, the results of the
corresponding models by Boss (2002) without magnetic braking
are shown as well.

 Figure 1 shows the initial equatorial density distribution
for the prolate cores with a 20:1 density contrast. Given
the stability of the initial models, evolutions consist of
the clouds oscillating about the initial configurations, 
waiting for sufficient time to elapse for ambipolar diffusion
to reduce the magnetic field support enough to allow collapse
to proceed, as in the previous magnetic cloud models (e.g.,
Boss 1997). Figure 2 shows the result for model P2BB,
which collapsed to form a quadruple protostellar system,
though in this case with an additional central density
maximum. Without magnetic braking, this core collapsed
to form a binary protostar system, so in this case magnetic
braking has led to an increased degree of fragmentation.
The off-axis clump evident in Figure 2 at 9 o'clock has
a maximum density of $2.5 \times 10^{-12}$ g cm$^{-3}$ and
an average temperature of 20 K. Considering regions with a
density at least 0.1 that of this maximum density, the clump's 
mass is 2.6 Jupiter masses, greater than the Jeans mass
at that density and temperature of 1.9 Jupiter masses.
The ratio of thermal to gravitational energy for the clump
is 0.49, showing that it is gravitationally bound. The
ratio of rotational to gravitational energy is 0.39,
so the clump is in rapid rotation. These fragment properties
are quite similar to those found in the Boss (2002) models.

 Model P2BA behaved in much the same way as P2BB, even with 
stronger magnetic braking ($f_{mb} = 0.001$ for P2BA compared to 
$f_{mb} = 0.0001$ for P2BB).

 Figure 3 shows the outcome of model P2BD, which started
with a lower initial angular velocity than model P2BB but was
otherwise identical. In this case, a well-defined binary
protostar system forms. A similar outcome resulted for
model P2BC with a higher degree of magnetic braking.
With even lower initial angular velocity than P2BC and
P2BD, model P2BE collapsed to form a binary-bar system
(Figure 4), i.e., a binary with its members connected
by a bar of rotating gas. In the case of model P2BE,
there are two density maxima in the bar near the center
of the system, possible evidence for further fragments,
though none of the density maxima are as well-defined
as those of the binary in Figure 3. In contrast, when
magnetic braking was neglected (Boss 2002), a core
identical to model P2BE fragmented into a well-defined
binary system, so in this case magnetic braking has
reduced the degree of fragmentation.

 The prolate core models with 100:1 initial density contrasts
all behaved in the same manner and formed single central
protostars embedded in bar-like structures. Figure 5 shows
the result for model P1BB, termed a single-bar. Figure 6
shows the equatorial temperature contours for this model,
emphasizing the highly non-uniform temperature distribution
that results from including radiative transfer effects
(e.g., Boss et al. 2000). The temperature field is strongest
in regions where the infalling gas is forming shock-like
density corrugations, yielding an x-shaped pattern in
the cloud's midplane. The formation of single-bars in
all five of the prolate 100:1 models compared to the formation of 
binaries in the corresponding Boss (2002) models shows that
when the initial cores are highly centrally condensed,
magnetic braking is able to frustrate fragmentation.

 Figure 7 shows the initial conditions for the oblate cores
with 20:1 initial density contrasts, while Figure 8 depicts
the outcome of model O2BB: a well-defined ring. While
the ring shows no particular tendency to fragment over the
time scale of the calculation, such a configuration is
expected to fragment eventually. With the exception of
model O2BA, which did not collapse significantly, all of
the oblate cores collapsed to form rings, for both 20:1
and 100:1 density contrasts, though the rings were more
pronounced for the cores with higher initial rotation rates,
such as O2BB in Figure 8. In comparison, the corresponding
Boss (2002) models formed a combination of rings or rings
which showed a tendency to fragment into quadruple systems.
Hence the oblate cloud models indicate that the inclusion
of magnetic braking had only a mild tendency to inhibit
their fragmentation.
 
 Tables 1 and 2 show that the prolate clouds tended to take
considerably longer to undergo collapse than the oblate clouds,
implying a considerably longer period in the pre-collapse,
quasi-equilibrium phase where thermal and magnetic support
dominate. As a result, the effect of magnetic braking through
the $f_{mb}$ approximation should be stronger in the models
which took the longest time to collapse, i.e., the models
in Table 1 with $\rho_0/\rho_R = 100:1$ and $t_f/t_{ff} \sim 9$.
Prolonged magnetic braking will tend to suppress rotationally-driven
fragmentation, consistent with the formation of single-bars
for the prolate models in Table 1 with $t_f/t_{ff} \sim 9$.
Model P2bb, for example, lost about 0.4\% of its total
angular momentum during its evolution to $t_f/t_{ff} = 4.683$, 
compared to a loss of about 1\% for model P1Bb within
$t_f/t_{ff} = 9.071$. Note though that for the oblate models 
in Table 2, no such effect is evident, perhaps because all of 
those models collapsed within $t_f/t_{ff} < 5$. In fact,
a comparision of the corresponding oblate models shows that
model O2bb lost about 0.4\% of its total angular momentum 
during its evolution to $t_f/t_{ff} = 2.456$, compared to a 
loss of about 1\% for model O1Bb within $t_f/t_{ff} = 4.636$.
Both models O2BB and O1BB collapse to form rings. Their percentage 
angular momentum losses are identical to those for models
P2BB and P1BB, which formed a quadruple and single-bar,
respectively, implying that the initial cloud shape and
degree of central concentration do have an important effect
on the fragmentation process of magnetic clouds.

\section{Conclusions}

 These pseudo-magnetohydrodynamics calculations have explored
the possibly deleterious effects of magnetic braking on
protostellar fragmentation, an issue explored by Hosking
\& Whitworth (2004). A degree of inhibition of fragmentation
caused by magnetic braking of both prolate and oblate, dense
cloud cores has been identified in these models through a
direct comparison with otherwise identical models calculated 
by Boss (2002) without magnetic braking. Nevertheless, a
cursory examination of the figures and tables reveals that
there is a still a large portion of initial conditions space
that appears to be permissive of fragmentation of magnetic
clouds when the approximate effects of magnetic pressure, 
tension, and braking are all included. The present 
models thus suggest that binary and multiple stars may well form 
from the collapse and fragmentation of magnetic, as well as 
non-magnetic, dense cloud cores, though perhaps not quite so readily.

 Given the approximate nature of the present calculations (Boss 2007),
it will be important to try to confirm these results with a true
magnetohydrodynamics code. This could be accomplished by adding
a numerical solution of the magnetic induction equation to the
Boss \& Myhill (1992) code. Alternatively, these calculations
could be repeated using publically available MHD codes, such as the 
FLASH adaptive mesh refinement code (e.g., Duffin \& Pudritz 2008),
though FLASH does not at present include Eddington approximation
radiative transfer, unlike the Boss \& Myhill (1992) code.
Attempting such true MHD calculations stands as a challenge for
future work on the question of protostellar collapse and 
fragmentation.

\acknowledgments

 The numerical calculations were performed on the Carnegie Alpha Cluster, 
the purchase of which was partially supported by the National Science 
Foundation under grant MRI-9976645. I thank Sandy Keiser for system 
management of the cluster, and the referee for numerous helpful
suggestions.

\clearpage
\begin{deluxetable}{ccccccc}
\tablecaption{Initial conditions and results for initially prolate 
cores. In this table and the following, $\rho_0/\rho_R$ is
the initial ratio of the central to the boundary density,
the units for the initial angular velocity $\Omega_i$ are 
rad s$^{-1}$, and the magnetic braking factor $f_{mb}$ is 
dimensionless. The final times $t_f$ are given 
in units of the initial free fall time $t_{ff} = 
(3\pi/32 G \rho_0)^{1/2} = 4.7 \times 10^4$ yr. 
Q denotes a core that collapses to form a quadruple system,
B denotes a binary outcome, BB a binary-bar, and SB a single-bar.
The results obtained by Boss (2002) for identical cores but 
without magnetic braking are shown as well.
\label{tbl-1}}
\tablehead{\colhead{\quad model } & 
\colhead{\quad $\rho_0/\rho_R$ } &
\colhead{\quad $\Omega_i$ \quad } & 
\colhead{\quad $f_{mb}$ \quad } &
\colhead{\quad $t_f$/$t_{ff}$ } & 
\colhead{\quad result \quad } &
\colhead{\quad Boss(2002) } }
\startdata

P2BA   &  20:1  & $1.0 \times 10^{-13}$  & 0.001  &  4.649 & Q & B \\

P2BB   &  20:1  & $1.0 \times 10^{-13}$  & 0.0001 &  4.683 & Q & B \\       

P2BC   &  20:1  & $3.2 \times 10^{-14}$  & 0.001  &  4.602 & B & B \\       

P2BD   &  20:1  & $3.2 \times 10^{-14}$  & 0.0001 &  4.603 & B & B \\       

P2BE   &  20:1  & $1.0 \times 10^{-14}$  & 0.001  &  4.833 & BB & B \\       

P1BA   & 100:1  & $1.0 \times 10^{-13}$  & 0.001  &  9.065 & SB & B \\       

P1BB   & 100:1  & $1.0 \times 10^{-13}$  & 0.0001 &  9.071 & SB & B \\       

P1BC   & 100:1  & $3.2 \times 10^{-14}$  & 0.001  &  8.942 & SB & B \\        

P1BD   & 100:1  & $3.2 \times 10^{-14}$  & 0.0001 &  8.945 & SB & B \\       

P1BE   & 100:1  & $1.0 \times 10^{-14}$  & 0.001  &  8.918 & SB & B \\       

\enddata
\end{deluxetable}
\clearpage

\clearpage
\begin{deluxetable}{ccccccc}
\tablecaption{Initial conditions and results for initially oblate
cores. R denotes a ring formed, while NC means no significant
collapse occurred. 
\label{tbl-2}}
\tablehead{\colhead{\quad model } & 
\colhead{\quad $\rho_0/\rho_R$ } &
\colhead{\quad $\Omega_i$ \quad } & 
\colhead{\quad $f_{mb}$ \quad } &
\colhead{\quad $t_f$/$t_{ff}$ } & 
\colhead{\quad result \quad } &
\colhead{\quad Boss(2002) } }
\startdata

O2BA   &  20:1  & $1.0 \times 10^{-13}$  & 0.001  & 4.565 & NC & R \\

O2BB   &  20:1  & $1.0 \times 10^{-13}$  & 0.0001 & 2.456 & R & R \\           

O2BC   &  20:1  & $3.2 \times 10^{-14}$  & 0.001  & 2.127 & R & Q \\  

O2BD   &  20:1  & $3.2 \times 10^{-14}$  & 0.0001 & 2.134 & R & Q \\  

O2BE   &  20:1  & $1.0 \times 10^{-14}$  & 0.001  & 2.109 & R & Q \\  

O1BA   & 100:1  & $1.0 \times 10^{-13}$  & 0.001  & 4.656 & R & R \\       

O1BB   & 100:1  & $1.0 \times 10^{-13}$  & 0.0001 & 4.636 & R & R \\       

O1BC   & 100:1  & $3.2 \times 10^{-14}$  & 0.001  & 4.521 & R & R \\   

O1BD   & 100:1  & $3.2 \times 10^{-14}$  & 0.0001 & 4.495 & R & R \\   

O1BE   & 100:1  & $1.0 \times 10^{-14}$  & 0.001  & 4.422 & R & Q \\

\enddata
\end{deluxetable}
\clearpage

\begin{figure}
\vspace{-2.0in}
\plotone{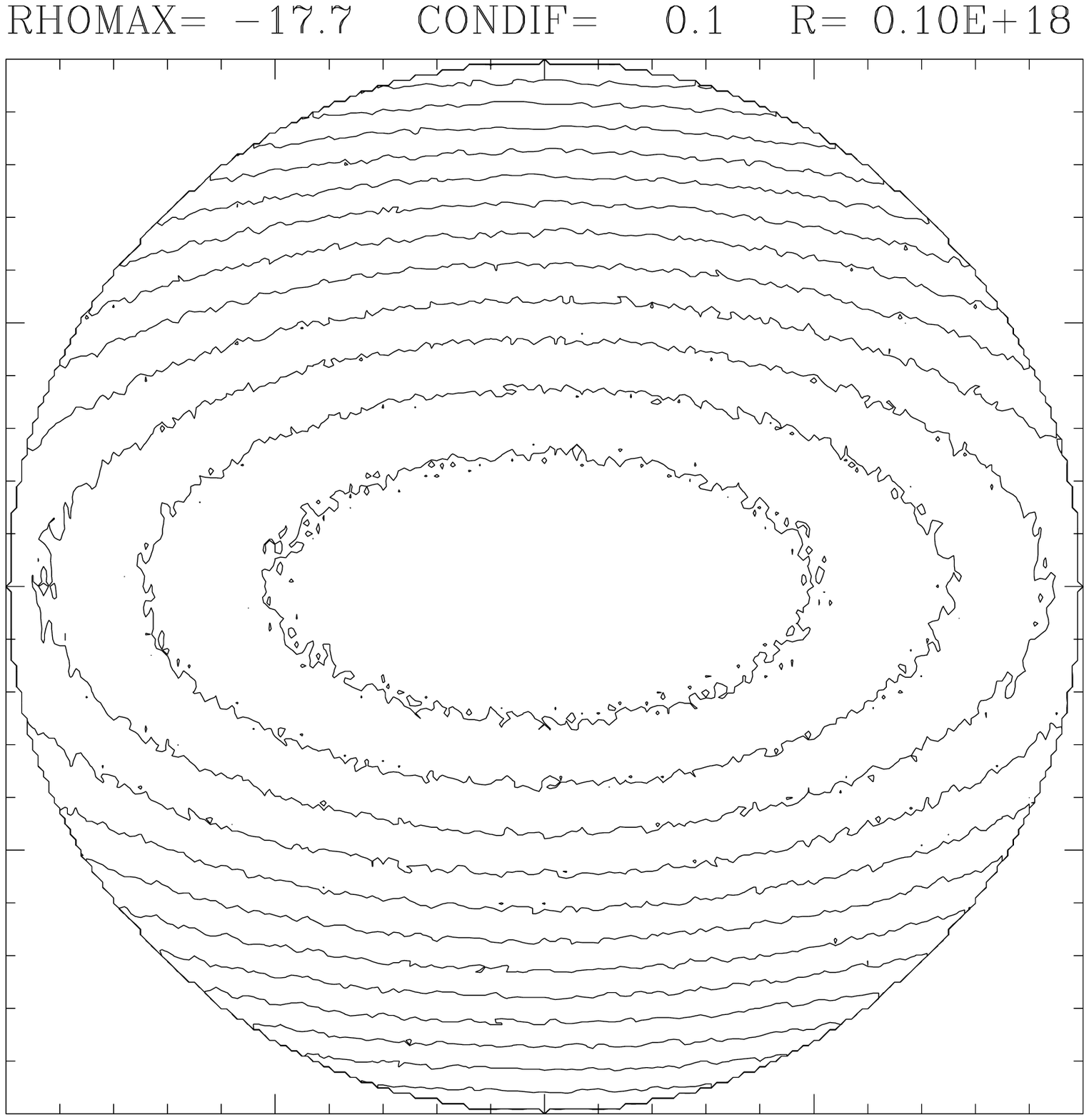}
\caption{Initial equatorial density contours for prolate core models 
with 20:1 density contrasts. Maximum density is $2.0 \times 10^{-18}$ 
g cm$^{-3}$. Contours represent changes by a factor of 1.3 in density. 
Region shown is $1.0 \times 10^{17}$ cm in radius.}
\end{figure}

\begin{figure}
\vspace{-2.0in}
\plotone{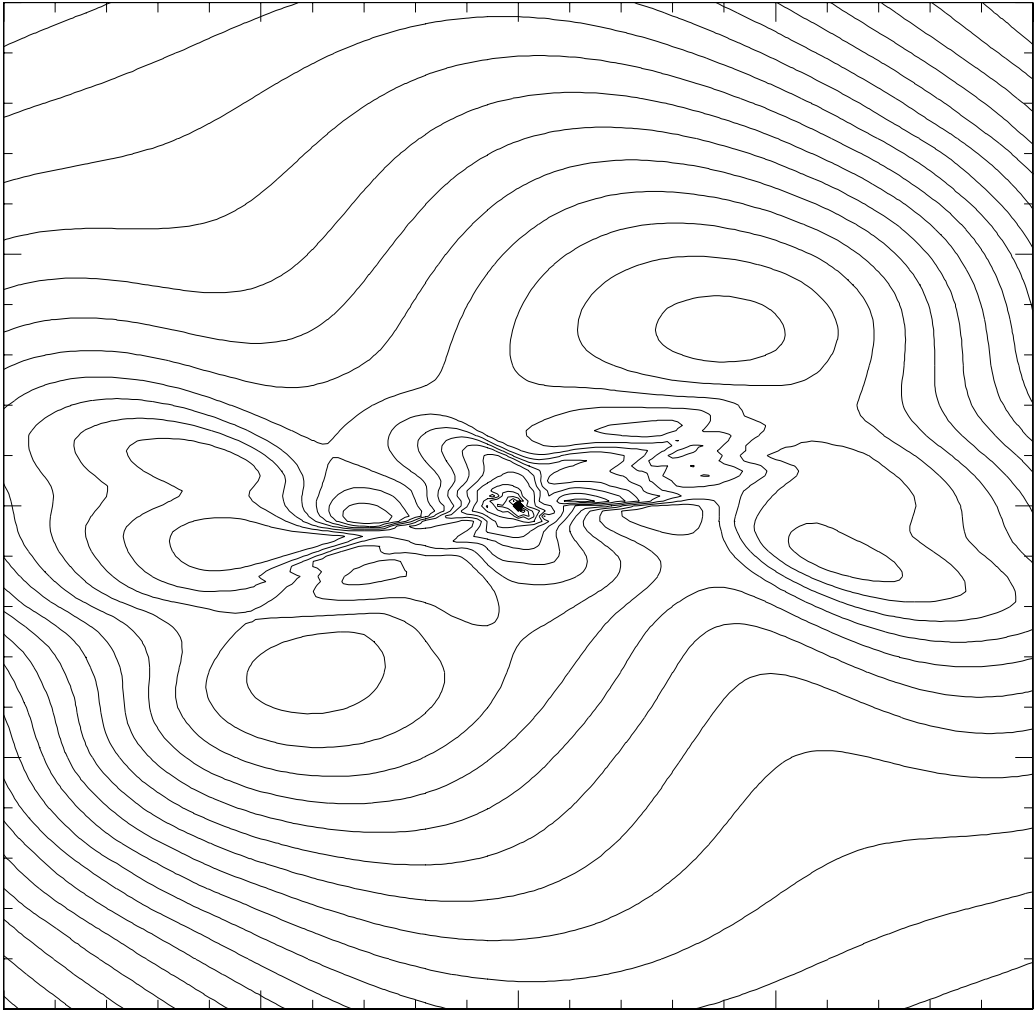}
\caption{Equatorial density contours for model P2BB at a time
of 4.683 $t_{ff}$. Maximum density is $1.0 \times 10^{-11}$ g cm$^{-3}$. 
Contours represent changes by a factor of 1.3 in density. Region shown is 
$4.0 \times 10^{14}$ cm in radius. A possible quintuple protostellar
system has formed: a central density maximum surrounded by
four density maxima.}
\end{figure}

\begin{figure}
\vspace{-2.0in}
\plotone{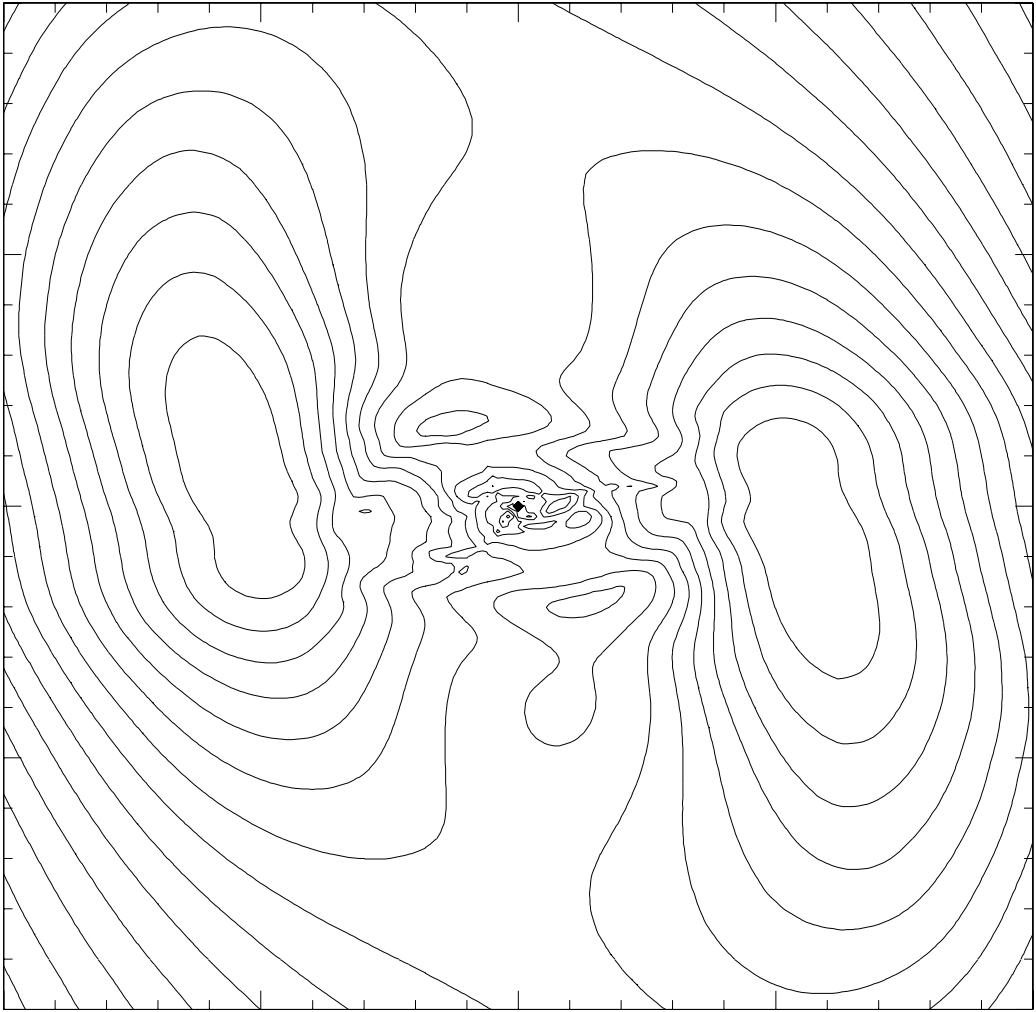}
\caption{Equatorial density contours for model P2BD at a time
of 4.603 $t_{ff}$. Maximum density is $1.3 \times 10^{-12}$ g cm$^{-3}$. 
Contours represent changes by a factor of 1.3 in density. Region shown is 
$4.0 \times 10^{14}$ cm in radius. A binary protostellar
system has formed.}
\end{figure}

\begin{figure}
\vspace{-2.0in}
\plotone{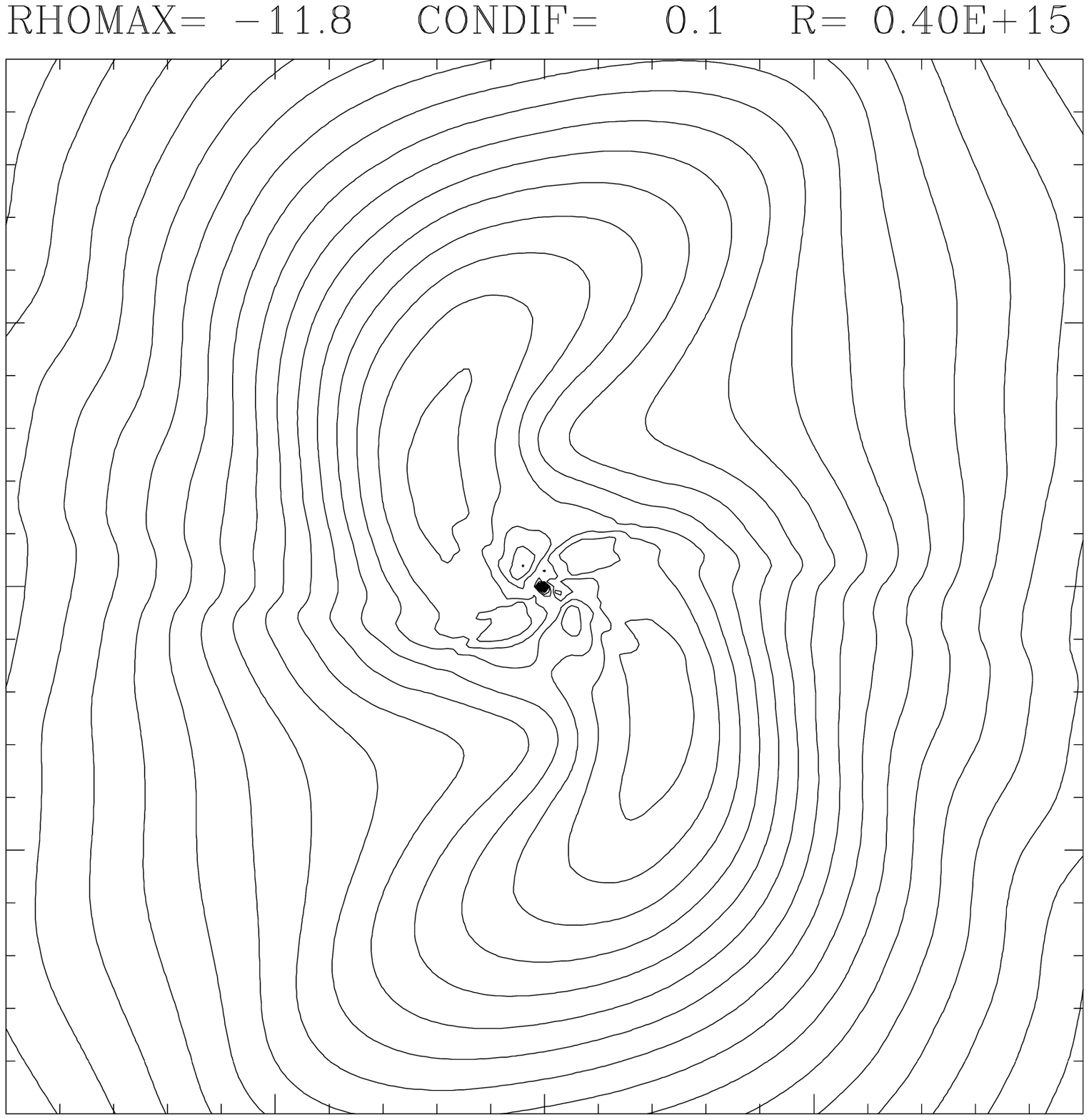}
\caption{Equatorial density contours for model P2BE at a time
of 4.833 $t_{ff}$. Maximum density is $1.6 \times 10^{-12}$ g cm$^{-3}$. 
Contours represent changes by a factor of 1.3 in density. Region shown is 
$4.0 \times 10^{14}$ cm in radius. A binary-bar system with four
different local density maxima has formed.}
\end{figure}

\begin{figure}
\vspace{-2.0in}
\plotone{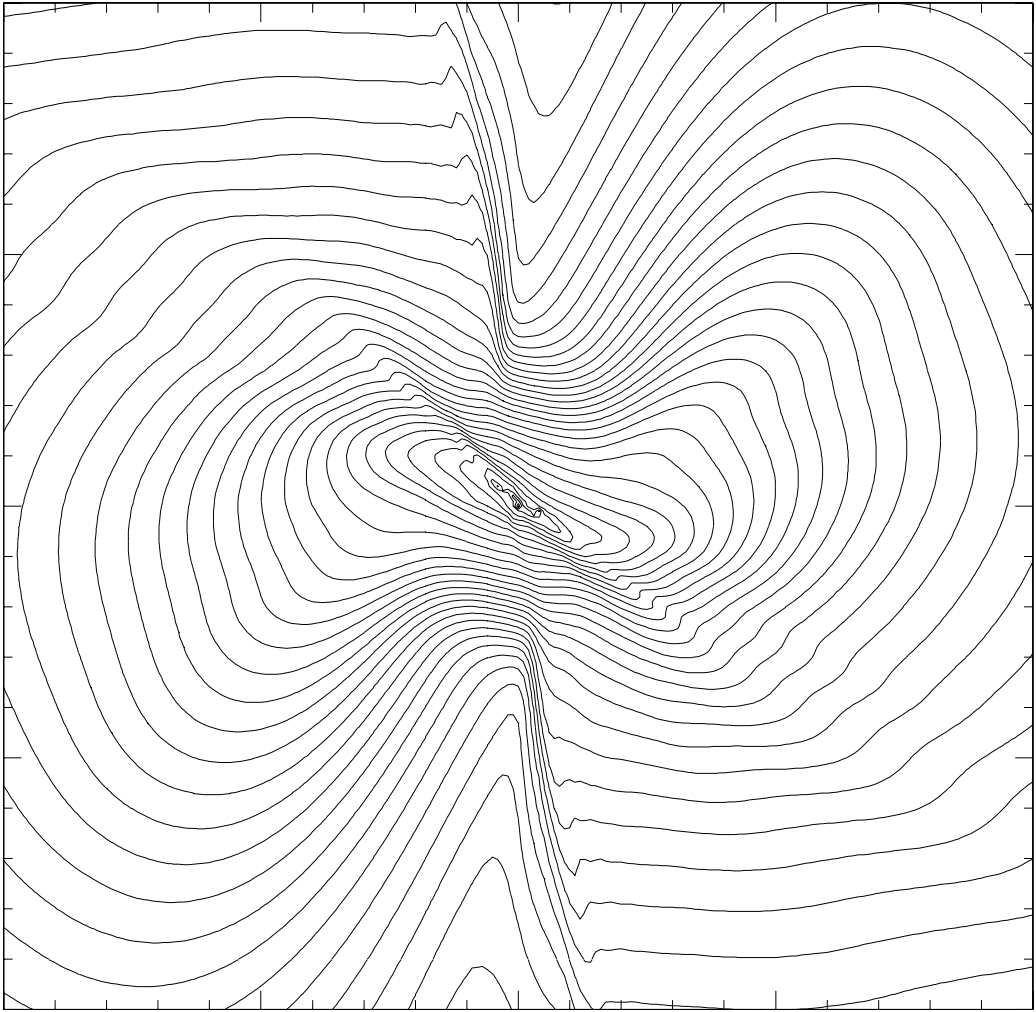}
\caption{Equatorial density contours for model P1BB at a time
of 9.069 $t_{ff}$. Maximum density is $6.3 \times 10^{-12}$ g cm$^{-3}$. 
Contours represent changes by a factor of 1.3 in density. Region shown is 
$4.0 \times 10^{14}$ cm in radius. A single-bar system has formed.}
\end{figure}

\begin{figure}
\vspace{-2.0in}
\plotone{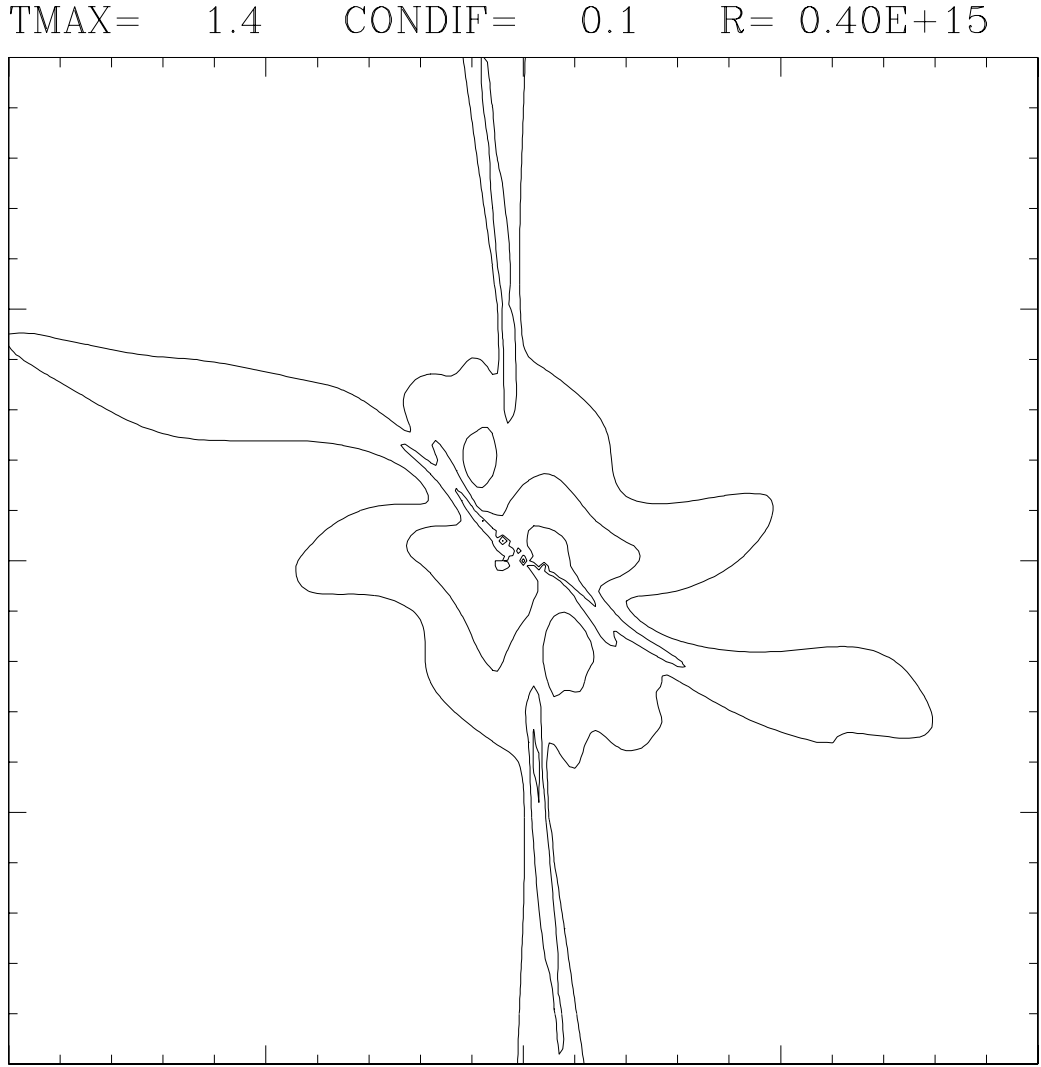}
\caption{Equatorial temperature contours for model P1BB at a time
of 9.069 $t_{ff}$. Maximum temperature is 25 K. Contours 
represent changes by a factor of 1.3 in temperature. Region shown is 
$4.0 \times 10^{14}$ cm in radius.} 
\end{figure}

\begin{figure}
\vspace{-2.0in}
\plotone{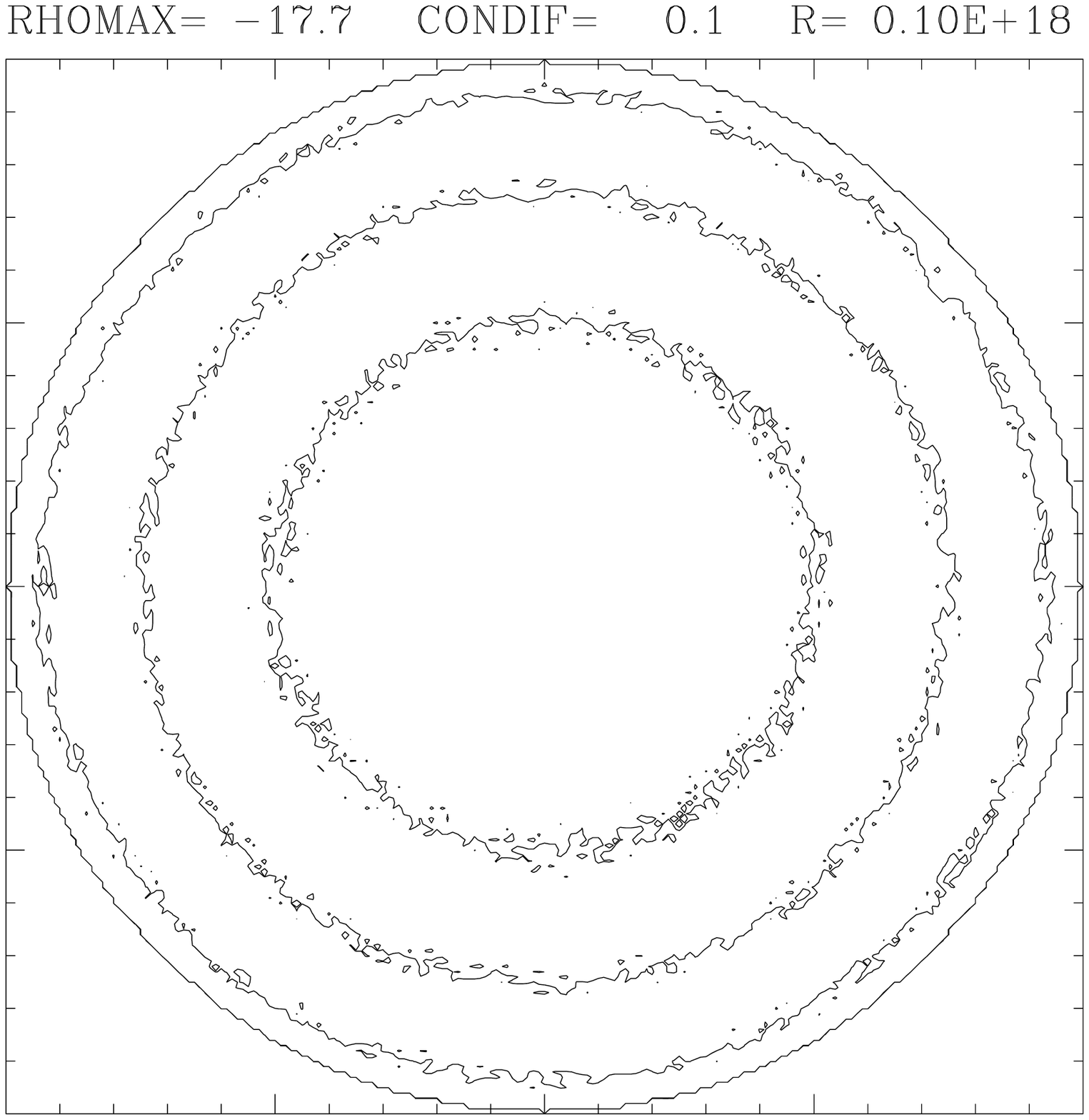}
\caption{Initial equatorial density contours for oblate core models 
with 20:1 density contrasts. Maximum density is $2.0 \times 10^{-18}$ 
g cm$^{-3}$. Contours represent changes by a factor of 1.3 in density. 
Region shown is $1.0 \times 10^{17}$ cm in radius.}
\end{figure}

\begin{figure}
\vspace{-2.0in}
\plotone{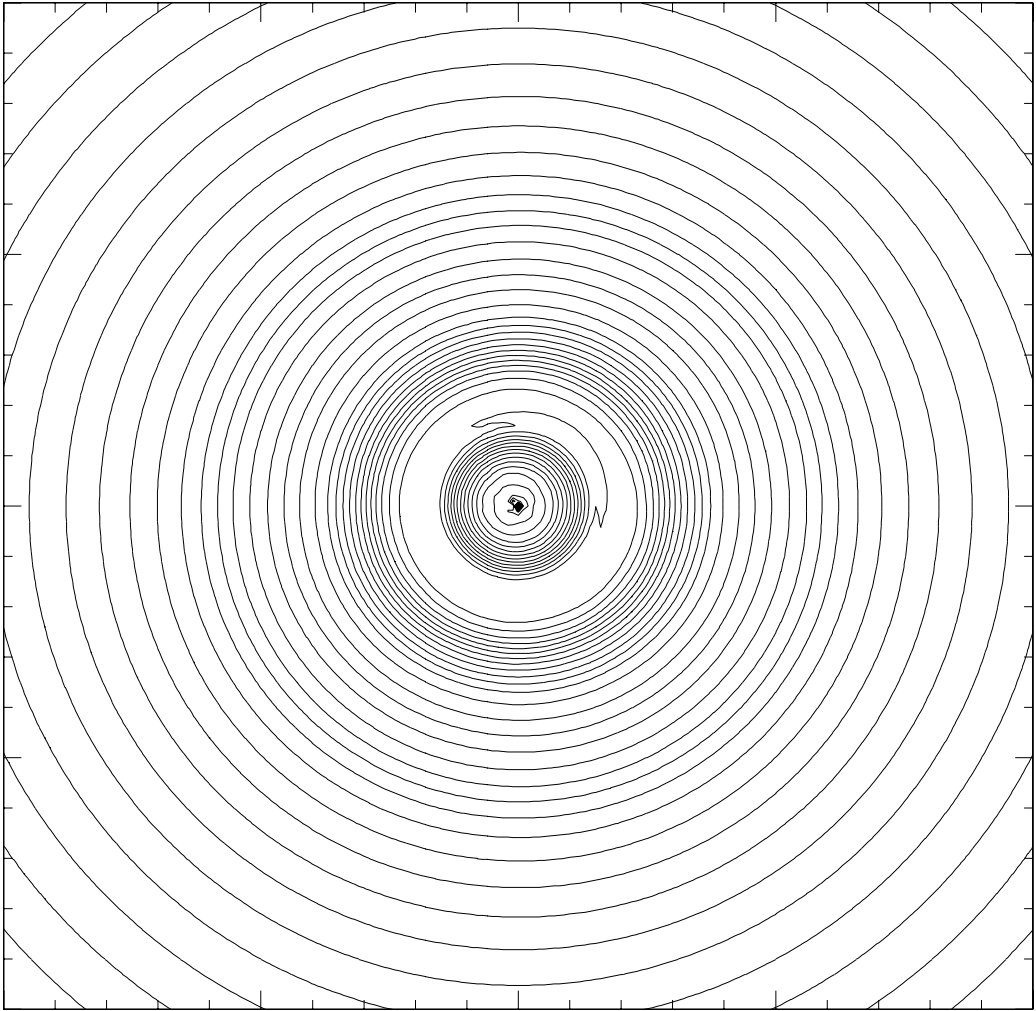}
\caption{Equatorial density contours for model O2BB at a time
of 2.456 $t_{ff}$. Maximum density is $2.0 \times 10^{-12}$ g cm$^{-3}$. 
Contours represent changes by a factor of 1.3 in density. Region shown is 
$2.3 \times 10^{15}$ cm in radius. A ring has formed: a strong density
minimum occurs at the center of the core.}
\end{figure}

\end{document}